\documentclass[aps,prd,preprint,showpacs,nofootinbib]{revtex4}
\usepackage{epsfig}
\begin{document}
\title{ Charged Higgs production at linear colliders in large extra dimensions}
\author{Qiang Li$^{a}$\footnote{\hspace{-0.1cm}
Electronics address: qliphy@pku.edu.cn}, Chong Sheng Li$^{a}$
\footnote{\hspace{-0.1cm} Electronics address: csli@pku.edu.cn},
Robert J. Oakes$^b$\footnote{\hspace{-0.1cm} Electronics address:
r-oakes@northwestern.edu}, and Li Lin
Yang$^{a}$\footnote{\hspace{-0.1cm} Electronics address:
llyang@pku.edu.cn} \\\ }
\address{\noindent$^a$ Department of Physics,
Peking University, Beijing 100871, China \\ $^b$ Department of
Physics and Astronomy, Northwestern University, Evanston, IL
60208-3112, USA}
\date{\today}

\begin{abstract}
In the Two-Higgs-Doublet Model(2HDM) with large extra
dimensions(LED), we study the contributions of virtual
Kaluza-Klein(KK) gravitons to 2HDM charged Higgs production,
especially in the two important production processes
$e^+e^-\rightarrow H^+H^-$ and $e^+e^-\rightarrow H^-t\bar{b}$, at
future linear colliders (LC). We find that KK graviton effects can
significantly modify these total cross sections and also their
differential cross sections compared to their respective 2HDM values
and, therefore, can be used to probe the effective scale $\Lambda_T$
up to several TeV. For example, at $\sqrt{s}=2$\,TeV, the cross
sections for $e^+e^-\rightarrow H^+H^-$ and $e^+e^-\rightarrow
H^-t\bar{b}$ in the 2HDM are 7.4fb for $m_{H^-}=150$\,GeV and
0.003fb for $m_{H^-}=1.1$\,TeV and $\tan\beta=40$ , while in LED
they are 12.1fb and 0.01fb, respectively, for $\Lambda_T=4$\,Tev.

\end{abstract}

 \maketitle
\section{Introduction}

The idea that quantum gravity can appear at the TeV energy scale
well below the Planck mass $M_{\rm p}\sim 1.2\times 10^{19}$GeV was
proposed in the 1990's\cite{ADD,RS,lyk,witt,hora,anto}. The large
extra dimensions(LED) model\cite{ADD} introduced by Arkani-Hamed,
Dimopoulos and Davli has attracted much attention. It has been
emphasized that the presence of large extra dimensions brings a new
solution to the hierarchy problem, which can take the place of other
mechanisms, for example, low-energy supersymmetry. However, it also
interesting to examine a scenario which combines new physics beyond
the Standard Model(SM) such as the 2HDM\cite{THDM} and LED. This new
possibility leads to different phenomenology than the usual LED
scenario,which we explore here.

In this extended LED scenario, as in the usual LED scheme, the total
space-time has $D=4+\delta$ dimensions. The SM and new particles
beyond the SM live in the usual $3+1-$dimensional space, while
gravity can propagate in the additional $\delta$-dimensional space,
which is assumed for simplicity to be compactified on the
$\delta$-dimensional torus $T^\delta$ with a common radius R. Then
the 4-dimensional Planck scale $M_{\rm{p}}$ is no longer the
relevant scale but is related to the fundamental scale $M_s$ as
follows\cite{ADD,csaki}:
\begin{eqnarray}\label{scale}
M^2_{\rm{p}}=M^{\delta+2}_s(2\pi R)^\delta,
\end{eqnarray}
where $M_s\sim {\rm TeV}$. According to Eq.~(\ref{scale}),
deviations from the usual Newtonian gravitational force law can be
expected at distances smaller than $R\sim
2.10^{-17}10^{\frac{32}{\delta}}\rm{cm}$\cite{csaki}. For
$\delta\geq 2$, LED is consistent with the current
experiments\cite{grav} since gravitational forces are not yet well
probed at distances less than sbout a millimeter (However for
$\delta=2$, there are constraints arising from, e.g., supernova
cooling, which require $M_s\geq 10-100 \rm{TeV}$ if $\delta=2$
\cite{csaki}).

2HDM LED can  be tested at future high energy colliders. In 2HDM
LED, just as in LED, there exist KK towers of massive spin-2
gravitons and scalars which can interact with the SM and beyond SM
fields. There are two classes of effects that can probe LED: real
graviton emission and virtual KK tower exchange.

At future linear colliders the search for one or more Higgs bosons
will be a central task. In the SM, the Higgs boson mass is a free
parameter with an upper bound of $m_H\leq600$
--- 800\,GeV \cite{massh}. Beyond the SM, the 2HDM is of particular theoretical
interest, in which the two complex Higgs doublets correspond to
eight scalar states and  spontanoues symmetry breaking leads to five
physical Higgs bosons: two neutral CP-even bosons $h^0$ and $H^0$,
one neutral CP-odd boson $A^0$, and two charged bosons $H^\pm$. The
$h^0$ is the lightest and is a SM-like Higgs boson especially in the
decoupling region ($m_{A^0}\gg m_{Z^0}$). The other four are not
SM-like  and their discovery, particularly the charged Higgs bosons,
would provide evidence for the 2HDM. If the $H^\pm$ bosons have mass
$m_{H^\pm}<m_t-m_b$, they will be produced mainly through the $t\to
bH^+$ decays of top quarks, which can be produced singly or in pairs
at an $e^+e^-$ LC \cite{pairp}. If there is sufficient center of
mass energy available, $\sqrt{s}>2m_{H^\pm}$, then charged Higgs
pair production, $e^+e^-\rightarrow H^+H^-$, will be the dominant
production mechanism\cite{pairp}. However, if
$m_{H^\pm}>\max(m_t-m_b,\sqrt s/2)$, then $H^\pm$ bosons can only be
produced singly. And $e^+e^-\rightarrow H^-t\bar{b}$\cite{htb,
htbqcd} is one of the most important single charged Higgs production
processes that can also be used to measure the relevant Yukawa
couplings. Therefore, for this process, we will focus on the regime
where $\sqrt s/2<m_{H^\pm}<\sqrt s-m_t-m_b$. For $\sqrt s=500$~GeV
(1000~GeV), this implies that $250<m_{H^\pm}< 320$~GeV ($500<
m_{H^\pm} < 820$~GeV).

There are several kinds of such 2HDM models. In the model called
type I, one Higgs doublet provides masses for both the up-type and
down-type quarks. In the type II model, one Higgs doublet gives
masses to the up-type quarks and the other one to the down-type
quarks. In the type III model, both doublets contribute to
generate the masses for up-type and down-type quarks. In this
paper, we will concentrate on the type II 2HDM(2HDM II), which are
is favored by the Minimal Supersymmetric Standard Model
(MSSM)\cite{mssm}. We will also briefly discussed the results in
the type I 2HDM.

In the following we consider the the contributions of virtual KK
gravitons to the 2HDM charged Higgs production, especially in the
two important production processes $e^+e^-\rightarrow H^+H^-$ and
$e^+e^-\rightarrow H^-t\bar{b}$ at future linear colliders. The
presentation is organized as follows: In Sect.~II we present the
calculations. In Sect.~III we give the numerical results and discuss
them. Sec.~IV contains a brief conclusion.

\section{Analytic Calulations}
In this section we derive the cross section  for 2HDM charged Higgs
production. The 2HDM diagrams, including the additional virtual KK
gravitons($G^{(\vec{n})}_{\mu \nu}$), which contribute to the
processes $e^+e^-\rightarrow H^+H^-$ and $e^+e^-\rightarrow
H^-t\bar{b}$ are presented in Figs.~1 and ~2, respectively.
\begin{figure}[h!]
\vspace{1.0cm} \centerline{\epsfig{file=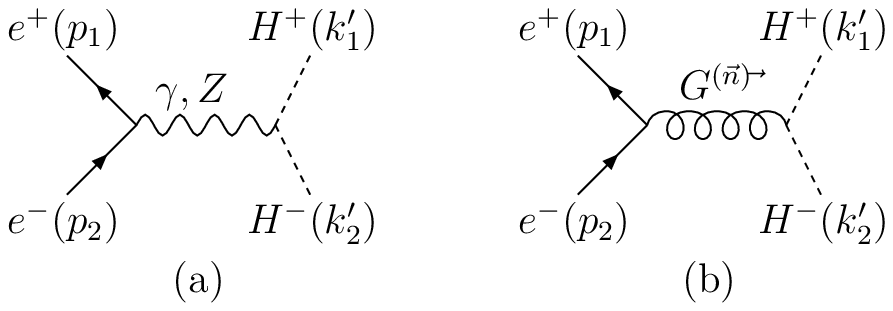, width=220pt}}
\caption[]{The 2HDM Feynman diagrams and graviton mediated diagrams
 for
$e^+e^-\rightarrow H^+H^-$ at the tree level.}
\end{figure}
\begin{figure}[h!]
\vspace{1.0cm} \centerline{\epsfig{file=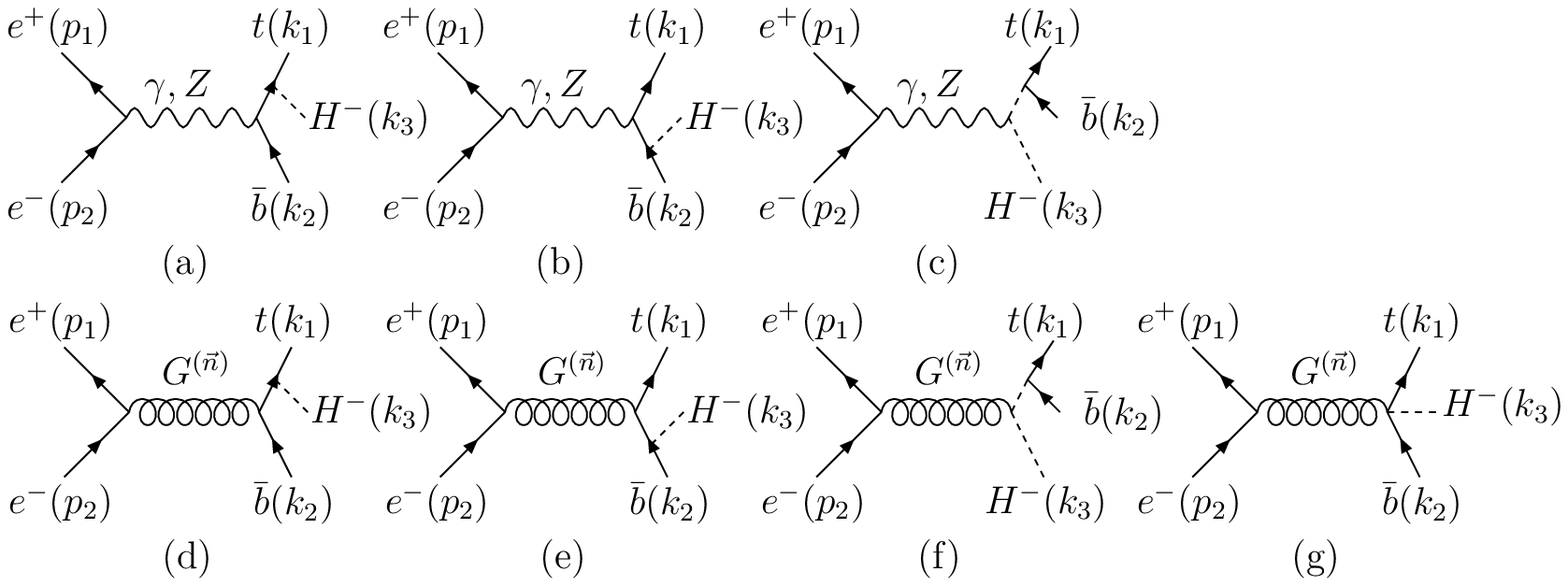, width=390pt}}
\caption[]{The 2HDM Feynman diagrams and graviton mediated diagrams
 for
$e^+e^-\rightarrow H^-t\bar{b}$ at the tree level.}
\end{figure}

In the four dimensional description the interaction Lagrangian
between the scattering fields and the KK gravitons ($
G^{(\vec{n})}_{\mu \nu} $) or KK scalars ($H^{(\vec{n})}$) is given
by \cite{Feynr1}
\begin{eqnarray}\label{IL}
 {\cal L}_{int} = - \frac{1}{{\bar M}_{p}}
\sum_{\vec{n}} \left( G^{(\vec{n})}_{\mu \nu} T^{\mu
\nu}-\frac{1}{3}\sqrt{\frac{3(n-1)}{n+2}} H^{(\vec{n})}
T^{\mu}_{\mu} \right) ,
\end{eqnarray}
 where
$\vec{n}=(n_1,n_2,..,n_\delta)$ with $n_i$'s being integers, ${\bar
M}_{p}=M_p/\sqrt{8 \pi}\sim 2.4\times 10^{18}$GeV is the reduced
four dimensional Planck scale, and $ T_{\mu \nu}$ is the
energy-momentum tensor of the scattering fields. The $n$-th KK mode
graviton and scaler masses squared are both characterized by
$m_{(\vec{n})}^2= |\vec{n}|^2/R^2$. Since the trace of the
energy-momentum tensor is proportional to the mass of the fields due
to the field equations, we neglect processes mediated by the KK
scalars in our study for the linear $e^+ e^-$ collider, and consider
only the processes mediated by the KK gravitons.

From Eq~.(\ref{IL}) we can derive the relevant Feynman rules to be
used in our calculations, which can be found in
Ref.~\cite{Feynr1,Feynr2}. The numerator of the graviton propagator
$P^{\mu\nu\alpha\beta}$ in the unitary gauge\cite{Feynr1}, is given
by:
\begin{eqnarray}\label{prop}
P^{\mu\nu\alpha\beta}=\frac{1}{2}\left(\eta^{\mu\alpha}\eta^{\nu\beta}
+\eta^{\mu\beta}\eta^{\nu\alpha}-
\frac{2}{3}\eta^{\mu\nu}\eta^{\alpha\beta}\right)
+\dots,\end{eqnarray}where $\eta^{\mu\nu}$ is the Minkowski metric.
The dots represent terms proportional to the graviton momentum
$q_{\mu}$, and since $q^{\mu}T_{\mu\nu}=0$, give a vanishing
contribution to the amplitude. Note that the the numerator of the
graviton propagator in Ref.~\cite{Feynr2} is twice Eq.~(\ref{prop})
and is the same as presented in Ref.~\cite{Feynr1}. However, the
coupling constant squared in Ref.~\cite{Feynr1}, i.e., $(\sqrt{8
\pi}/M_p)^2$ in Eq.~(\ref{IL}) is twice  the one in
Ref.~\cite{Feynr2}. Hence the results in the two references are
consistent and we have been careful of this issue in our
calculations.

We can write the amplitudes for the Feynman diagrams shown in Figs.1
and 2 as follows:
\begin{eqnarray}\label{ampl}
{\cal M}_{1a}&=& \frac{ie^2}{s}{\bar
v(p_1)}(\not{\!k'}_2-\not{\!k'}_1)u(p_2)
+\frac{ie^2(c^2_w-s^2_w)}{2(s-m_z^2)c_w^2s_w^2}{\bar
v(p_1)}(\not{\!k'}_2-\not{\!k'}_1)\big [(\frac{1}{2}-s^2_w)P_L
-s^2_wP_R\big]u(p_2)  \nonumber \\
{\cal M}_{1b}&=&  \frac{i{\cal
G}P^{\mu\nu\alpha\beta}}{4}{\chi}^{a}_{\alpha\beta}{\bar v(p_1)}C_{\mu\nu} u(p_2)\nonumber \\
%-=-=-=-=-=-=-=-=-=-=-=-=-=-=-=-=-=-=-=-=-=-=-=-=-=-=-=-=-=-=-=-=-=-=-=-=
{\cal M}_{2a}&=&
\frac{-ie^2g}{3\sqrt{2}m_ws[(k_1+k_3)^2-m_b^2]}{\bar
v(p_1)}\gamma_{\mu}u(p_2){\bar u(k_1,m_t)}\bigg [ m_b\tan\beta
P_R+m_t\cot\beta P_L \bigg ]
\nonumber\\
\hspace{1.0cm}& &(\not{\!k}_1+\not{\!k}_3+m_b)\gamma^{\mu}v(k_2,m_b)
+\nonumber\\
\hspace{1.0cm}&
&\frac{-ie^2g}{\sqrt{2}m_w(s-m^2_z)c^2_ws^2_w[(k_1+k_3)^2-m_b^2]}{\bar
v(p_1)}\gamma_{\mu}\big [(\frac{1}{2}-s^2_w)P_L
-s^2_wP_R\big]u(p_2).{\bar u(k_1,m_t)}\nonumber\\
\hspace{1.0cm}& &\bigg [ m_b\tan\beta P_R+m_t\cot\beta P_L \bigg
](\not{\!k}_1+\not{\!k}_3+m_b)\gamma^{\mu}[(\frac{1}{2}-\frac{1}{3}s^2_w)P_L
-\frac{1}{3}s^2_wP_R\big]
v(k_2,m_b)\nonumber\\
%-=-=-=-=-=-=-=-=-=-=-=-=-=-=-=-=-=-=-=-=-=-=-=-=-=-=-=-=-=-=-=-=-=-=-=-=-=-=-=-=-=-=-=-=-=-=-=-=
%-=-=-=-=-=-=-=-=-=-=-=-=-=-=-=-=-=-=-=-=-=-=-=-=-=-=-=-=-=-=-=-=-=-=-=-=
{\cal M}_{2b}&=&
\frac{-2ie^2g}{3\sqrt{2}m_ws[(k_2+k_3)^2-m_t^2]}{\bar
v(p_1)}\gamma_{\mu}u(p_2){\bar
u(k_1,m_t)}\gamma^{\mu}(\not{\!k}_2+\not{\!k}_3-m_t)
\nonumber\\
\hspace{1.0cm}& &\bigg [ m_b\tan\beta P_R+m_t\cot\beta P_L \bigg
]v(k_2,m_b)
+\nonumber\\
\hspace{1.0cm}&
&\frac{ie^2g}{\sqrt{2}m_w(s-m^2_z)c^2_ws^2_w[(k_2+k_3)^2-m_t^2]}{\bar
v(p_1)}\gamma_{\mu}\big [(\frac{1}{2}-s^2_w)P_L
-s^2_wP_R\big]u(p_2).{\bar u(k_1,m_t)}\nonumber\\
\hspace{1.0cm}& &\gamma^{\mu}[-(\frac{1}{2}-\frac{2}{3}s^2_w)P_L
+\frac{2}{3}s^2_wP_R\big](\not{\!k}_2+\not{\!k}_3-m_t)\bigg [
m_b\tan\beta P_R+m_t\cot\beta P_L \bigg ]v(k_2,m_b)
\nonumber\\
%-=-=-=-=-=-=-=-=-=-=-=-=-=-=-=-=-=-=-=-=-=-=-=-=-=-=-=-=-=-=-=-=-=-=-=-=-=-=-=-=-=-=-=-=-=-=-=-=
%-=-=-=-=-=-=-=-=-=-=-=-=-=-=-=-=-=-=-=-=-=-=-=-=-=-=-=-=-=-=-=-=-=-=-=-=
{\cal M}_{2c}&=&
\frac{-ie^2g}{\sqrt{2}m_ws[(k_1+k_2)^2-m_{H^-}^2]}{\bar
v(p_1)}(2\not{\!k}_3-\not{\!p}_1-\not{\!p}_2)u(p_2)\nonumber\\
\hspace{1.0cm}& &{\bar u(k_1,m_t)}\bigg [ m_b\tan\beta
P_R+m_t\cot\beta P_L \bigg ] v(k_2,m_b)
+\nonumber\\
\hspace{1.0cm}&
&\frac{-ie^2g(c^2_w-s^2_w)}{2\sqrt{2}m_w(s-m^2_z)c^2_ws^2_w[(k_1+k_2)^2-m_{H^-}^2]}{\bar
v(p_1)}(2\not{\!k}_3-\not{\!p}_1-\not{\!p}_2)[(\frac{1}{2}-s^2_w)P_L
-s^2_wP_R\big]u(p_2)\nonumber\\
\hspace{1.0cm}& &{\bar u(k_1,m_t)}\bigg [ m_b\tan\beta
P_R+m_t\cot\beta P_L \bigg ] v(k_2,m_b)\nonumber
\end{eqnarray}
%-=-=-=-=-=-=-=-=-=-=-=-=-=-=-=-=-=-=-=-=-=-=-=-=-=-=-=-=-=-=-=-=-=-=-=-=
\begin{eqnarray}
{\cal M}_{2d}&=&  \frac{-ig{\cal
G}P^{\mu\nu\alpha\beta}}{16\sqrt{2}m_w[(k_1+k_3)^2-m_{b}^2]} {\bar
v(p_1)}C_{\mu\nu}u(p_2)\nonumber\\
\hspace{1.0cm}& &{\bar u(k_1,m_t)}\bigg [ m_b\tan\beta
P_R+m_t\cot\beta P_L \bigg ](\not{\!k}_1+\not{\!k}_3+m_b){\chi}^{b}_{\alpha\beta}v(k_2,m_b)\nonumber \\
%-=-=-=-=-=-=-=-=-=-=-=-=-=-=-=-=-=-=-=-=-=-=-=-=-=-=-=-=-=-=-=-=-=-=-=-=
{\cal M}_{2e}&=&  \frac{ig{\cal
G}P^{\mu\nu\alpha\beta}}{16\sqrt{2}m_w[(k_2+k_3)^2-m_{t}^2]} {\bar
v(p_1)}C_{\mu\nu}u(p_2)\nonumber\\
\hspace{1.0cm}& &{\bar
u(k_1,m_t)}{\chi}^{c}_{\alpha\beta}(\not{\!k}_2+\not{\!k}_3-m_t)\bigg
[ m_b\tan\beta P_R+m_t\cot\beta P_L \bigg ]v(k_2,m_b)\nonumber\\
%-=-=-=-=-=-=-=-=-=-=-=-=-=-=-=-=-=-=-=-=-=-=-=-=-=-=-=-=-=-=-=-=-=-=-=-=
{\cal M}_{2f}&=&  \frac{-ig{\cal
G}P^{\mu\nu\alpha\beta}{\chi}^{d}_{\alpha\beta}}{4\sqrt{2}m_w[(k_1+k_2)^2-m_{H^-}^2]}
{\bar
v(p_1)}C_{\mu\nu}u(p_2)\nonumber\\
\hspace{1.0cm}& &{\bar u(k_1,m_t)}\bigg [ m_b\tan\beta
P_R+m_t\cot\beta P_L \bigg ]v(k_2,m_b)\nonumber\\
%-=-=-=-=-=-=-=-=-=-=-=-=-=-=-=-=-=-=-=-=-=-=-=-=-=-=-=-=-=-=-=-=-=-=-=-=
{\cal M}_{2g}&=& 0
\end{eqnarray}
with
\begin{eqnarray}\label{coe}
{\cal G} & =&
\frac{-1}{\bar{M}^2_p}\sum_{\vec{n}}\frac{1}{s-m^2_{\vec{n}}}
\\
C_{\mu\nu} & = &
    \big[\gamma_\mu \big(p_2 - p_1)_{\nu} + (\mu \leftrightarrow \nu) \big] \\
%-------------------------------------------------------------------------------
{\chi}^{a}_{\alpha\beta} & = &m_{H^-}^2 \eta_{\alpha\beta} -
(k'_1)^\mu (k'_2)^\nu \left[ \eta_{\mu \alpha} \eta_{\nu \beta}
                             + \eta_{\nu \alpha} \eta_{\mu \beta}
                             - \eta_{\mu \nu} \eta_{\alpha \beta} \right]\\
%---------------------------------------------------------------------------------
{\chi}^{b}_{\alpha\beta} & = &
       \left[ \gamma_{\alpha} \big(k_1+k_3-k_2)_{\beta}
              - \eta_{\alpha\beta}
               \big( k_1{\!\!\!\!/}+k_3{\!\!\!\!/}-k_2{\!\!\!\!/}- 2 m_b\big)
                     \right]
       + (\beta \leftrightarrow \alpha ) \\
%-----------------------------------------------------------------------
{\chi}^{c}_{\alpha\beta} & = &
       \left[ \gamma_{\alpha} \big(k_1-k_3-k_2)_{\beta}
              - \eta_{\alpha\beta}
               \big( k_1{\!\!\!\!/}-k_2{\!\!\!\!/}-k_3{\!\!\!\!/}- 2 m_t\big)
                     \right]
       + (\beta \leftrightarrow \alpha ) \\
%-----------------------------------------------------------------------
{\chi}^{d}_{\alpha\beta} & = &m_{H^-}^2 \eta_{\alpha\beta} - k_3^\mu
(k_1+k_2)^\nu \left[ \eta_{\mu \alpha} \eta_{\nu \beta}
                             + \eta_{\nu \alpha} \eta_{\mu \beta}
                             - \eta_{\mu \nu} \eta_{\alpha \beta}
                             \right],
\end{eqnarray}
where $s \equiv (p_1 + p_2)^2 $, $P_L=(1-\gamma_5)/2$,
$P_R=(1+\gamma_5)/2$, $s_w=\sin_w$, $c_w=\cos_w$ and $\tan\beta$ is
the ratio of the two vacuum expectation values in the 2HDM. Note
that the contribution from Feynman diagram Fig.2(g) vanishes since
the trace of the graviton appears in this diagram and, therefore, in
the limit of vanishing electron mass, this contribution also
vanishes\cite{vanish}.

${\cal G}$ in Eq.~(\ref{coe}) represents the summation of the KK
excitation propagators. If the summation over the infinite tower of
the KK modes is performed, one will encounter ultraviolet
divergences. This happens because LED is an effective theory, which
is only  valid below an effective energy scale. In the following
calculations we naively introduce an ultraviolet cutoff for the
highest KK modes and replace the summation by\cite{Feynr1}
\begin{eqnarray}\label{cut}
\frac{4\pi}{\Lambda^4_T}=\frac{-1}{\bar{M}^2_p}\sum_{\vec{n}}\frac{1}{s-m^2_{\vec{n}}},
\end{eqnarray}
where $\Lambda_T$ is a cutoff scale naturally being of the order of
the fundamental  scale $M_s$.

Finally, the cross sections for the charged Higgs production
processes in LED following from the amplitudes are:
\begin{eqnarray}\label{sigma}
&&\sigma(e^+e^-\rightarrow H^+H^-)=\frac{1}{2s}\int d\Phi_2
\frac{1}{4}|M_{1a}+M_{1b}|^2\\
&&\sigma(e^+e^-\rightarrow H^-t\bar{b})=\frac{1}{2s}\int d\Phi_3
\frac{3}{4}|M_{2a}+M_{2b}+M_{2c}+M_{2d}+M_{2e}+M_{2f}+M_{2g}|^2,
\end{eqnarray}
and the final state phase space hypercube elements are defined as
\begin{eqnarray}
&&{\rm d} \Phi_2 =
   \left( \prod_{i=1}^2 \frac{{\rm d}^3 \vec{k'}_i}{(2\pi)^3 2(k'_i)^0} \right)\, (2\pi)^4
   \delta\Biggl(p_1+p_2-\sum_{j=1}^2 k'_j\Biggr).
\\
&&{\rm d} \Phi_3 =
   \left( \prod_{i=1}^3 \frac{{\rm d}^3 \vec{k}_i}{(2\pi)^3 2k_i^0} \right)\, (2\pi)^4
   \delta\Biggl(p_1+p_2-\sum_{j=1}^3 k_j\Biggr).
\end{eqnarray}
\section{numerical results}
In the numerical calculations, we used the following set of SM
parameters\cite{SM}:
\begin{eqnarray}
 \alpha_{ew}(m_W)=1/128,  m_W=80.419 {\rm GeV}, m_t=178
{\rm GeV},  m_Z=91.1882{\rm GeV}.
\end{eqnarray}
For the Yukawa coupling of the bottom quark at the $H^-t\bar{b}$
vertex, as suggested by Ref.\cite{htbqcd}, we used the top quark
pole mass and the QCD improved running mass $m_b(Q)$ with
$m_b(m_b)=4.25$\,GeV, which were evaluated using the NLO formula
\cite{runningmb} as follows:
\begin{equation}
m_b(Q)=U_6(Q,m_t)U_5(m_t,m_b)m_b(m_b) \, ,
\end{equation}
Here the evolution factor $U_f$ is
\begin{eqnarray}
U_f(Q_2,Q_1)=\bigg(\frac{\alpha_s(Q_2)}{\alpha_s(Q_1)}\bigg)^{d^{(f)}}
\bigg[1+\frac{\alpha_s(Q_1)-\alpha_s(Q_2)}{4\pi}J^{(f)}\bigg], \nonumber \\
d^{(f)}=\frac{12}{33-2f}, \hspace{1.0cm}
J^{(f)}=-\frac{8982-504f+40f^2}{3(33-2f)^2} \, ,
\end{eqnarray}
and $f$ is the number of active light quarks. For the energy scale
$Q$, we chose $Q=\sqrt[3]{m_tm_bm_{H^-}}$ as in Ref.~\cite{htbqcd}
for $e^+e^-\rightarrow H^-t\bar{b}$.

\subsection{$e^+e^-\rightarrow H^+H^-$}

In Fig.~3 we show the dependence of the the cross sections in 2HDM
II and LED for the process $e^+e^-\rightarrow H^+H^-$ on $m_{H^-}$
assuming  $\sqrt{s}=1000$\,GeV. The dashed lines represent the
2HDM II results and the solid lines represent the cross section in
2HDM II LED for different values of $\Lambda_T$. From the figures
one can see that when $m_{H^-}$ is small($<400$\,GeV) the LED
effects can be very large up to $\Lambda_T=2$\,TeV. For example,
when $\sqrt{s}=1$\,TeV and $m_{H^-}=200$\,GeV the cross section in
the 2HDM is 23.8fb, while for 2HDM II LED it is 152.4fb for
$\Lambda_T=1.5$\,TeV and 36.6fb for $\Lambda_T=2$\,TeV.

In Fig.~4, the cross section in the 2HDM II and LED for the
process $e^+e^-\rightarrow H^+H^-$  are plotted as functions of
$\sqrt{s}$ assuming $m_{H^-}=200$\,GeV. From this figure we see
that the effects of virtual graviton exchange can substantially
modify the $e^+e^-\rightarrow H^+H^-$ cross section compared to
its 2HDM II value, especially at large $\sqrt{s}$($>4$\,TeV),
enabling LED up to $\Lambda_T=8$\,TeV to be probed.

In Fig.~5 we show the dependence of the cross sections in the 2HDM
II and LED for the process $e^+e^-\rightarrow H^+H^-$ on the
effective energy scale $\Lambda_T$ assuming $m_{H^-}=150$\,GeV and
$\sqrt{s}=1000$\,GeV, 2000\,GeV, and 3000\,GeV. As expected, when
$\sqrt{s}$ is fixed the cross section in LED tends to the value in
the 2HDM II when $\Lambda_T$ tends becomes very large. In
addition, the figure also shows that for large $\sqrt{s}$ and
small $\Lambda_T$ the LED effects are very large. For example, for
$\sqrt{s}$=3\,TeV and $\Lambda_T$=4\,TeV the cross section in the
2HDM II is 3.4fb while the value in LED is 58.7fb.

In Fig.~6, we present the differential cross section
$d\sigma/dcos\theta$ in the 2HDM II and LED for the process
$e^+e^-\rightarrow H^+H^-$ as a function of $\cos\theta$, where
the scattering angle $\theta$ is the angle between $H^-$ and the
incoming positron assuming $m_{H^-}=200$\,GeV,
$\sqrt{s}=1000$\,GeV, and $\Lambda_T=2000$\,GeV. Both differential
cross sections in the 2HDM II and LED vanish at $\cos\theta=\pm
1$. However, their shapes are different. Note that the LED cross
section is not symmetric under $\cos\theta\leftrightarrow
-\cos\theta$ while the 2HDM II one is symmetric. We further have
plotted the different contributions to the LED cross section and
find the asymmetry comes from the interference of the LED and 2HDM
II amplitudes, which is in proportion to
$(t-u)(tu-m^4_{H^-})/\Lambda^2_T$( $t$ and $u$ are Mandelstam
variables) and thus to $\cos\theta\sin^2\theta$ . Moreover, the
LED contribution has peaks in both the forward and backward
regions and also vanishes at three points: $\cos\theta=0$ and $\pm
1$, as found in Ref.~\cite{diff}, since it is proportional to
$(t-u)^2(tu-m^4_{H^-})/\Lambda^4_T$ and thus to
$\cos^2\theta\sin^2\theta$.

\subsection{$e^+e^-\rightarrow H^-t\bar{b}$}

In Fig.~7, we show the dependence of the the cross section in the
2HDM II and LED for the process $e^+e^-\rightarrow H^-t\bar{b}$ on
$m_{H^-}$  assuming $\sqrt{s}=1000$\,GeV for $\tan\beta=40$. The
dashed lines represent the 2HDM II results and the solid lines
represent the cross sections for LED with different values of
$\Lambda_T$. This figure shows that when $\Lambda_T$ is
small($<2.5$\,TeV) and $m_{H^-}$ is also small($<600$\,GeV), the
LED effects are very large. For example, when $\tan\beta=40$ and
$m_{H^-}=550$\,GeV the cross section in the 2HDM II is 0.017fb.

In Fig.~8, we present the dependence of the cross section in the
2HDM II and for LED for the process $e^+e^-\rightarrow
H^-t\bar{b}$ on $\tan\beta$ assuming $\sqrt{s}=1000$\,GeV and
$m_{H^-}=520$\,GeV. Observe that the cross section in both the
2HDM II and LED exhibit minima close to
$\tan\beta\approx\sqrt{m_t/m_b}\approx6$, since the average
strength of the $t\overline{b}H^-$ coupling, which is proportional
to $\sqrt{m_t^2\cot^2\beta+m_b^2\tan^2\beta}$, is then minimal
\cite{minm}. From these figures one also sees that when
$\Lambda_T=2$ is small($<2.5$\,TeV) and $m_{H^-}$ is also
small($<600$\,GeV) the LED effects are large for
$\sqrt{s}=1$\,TeV.

In Fig.~9 the cross sections in the 2HDM II and LED for the
process $e^+e^-\rightarrow H^-t\bar{b}$  are plotted as functions
of $\sqrt{s}$, assuming $m_{H^-}=800$\,GeV and $\tan\beta=40$.
Since we are interested in the case where $\sqrt
s/2<m_{H^\pm}<\sqrt s-m_t-m_b$,  $\sqrt{s}$ should be in the range
from about 983\,GeV to 1600\,GeV. From this figure we see that, as
in the case of $e^+e^-\rightarrow H^+H^-$, the effects of virtual
graviton exchange can significantly modify the $e^+e^-\rightarrow
H^-t\bar{b}$ cross section compared to its 2HDM II value,
especially at large $\sqrt{s}$($>1.5$\,TeV), where LED up to about
$\Lambda_T=3.5$\,TeV can be probed.

In Fig.~10 we show the dependence of the cross section in the 2HDM
II and LED for the process $e^+e^-\rightarrow H^-t\bar{b}$ on the
effective energy scale $\Lambda_T$ assuming $m_{H^-}=260, 550, and
1100$\,GeV,  and $\sqrt{s}=500, 1000$, and $2000$\,GeV,
respectively. As expected, when $\sqrt{s}$ is fixed the LED cross
section tends to the value in the 2HDM II when $\Lambda_T$ becomes
very large. This figure also shows that for large $\sqrt{s}$ and
small $\Lambda_T$ the LED effects can be very large. For example,
when $\sqrt{s}$=2\,TeV and $\Lambda_T$=4\,TeV the cross section in
the 2HDM II is 0.006fb, while in LED it is 0.013fb.

In Fig.~11 we show the differential cross sections
$d\sigma/dP_T(H^-)$ in the 2HDM II and LED for the process
$e^+e^-\rightarrow H^-t\bar{b}$ as functions of the transverse
momentum of the charged Higgs boson $P_T$ assuming
$m_{H^-}=520$\,GeV, $\sqrt{s}=1000$\,GeV, and
$\Lambda_T=1500$\,GeV, $2000$\,GeV, $2500$\,GeV and $3000$\,GeV.
The shape of the differential cross sections in the 2HDM II and
for LED are slightly different. The LED effects generally enhance
the differential cross sections, especially when $\Lambda_T$ is
small.

We briefly note the relevant results in type I 2HDM. The only
change is in the fermionic couplings. Thus, the 2HDM and LED
results for $e^+e^-\rightarrow H^+H^-$ remain the same. However,
the 2HDM results for $e^+e^-\rightarrow H^-t\bar{b}$ in the type I
model are different compared with the type II model. For example,
the results in type I model will decrease with increasing
$\tan\beta$, since the $H^-t\bar{b}$ coupling is proportional to
$\cot\beta$ in the type I model. However, the LED effects will be
similar: large effects when $\sqrt{s}$ is large and $\Lambda_T$ is
small.

Finally, we briefly discuss the experimental sensitivity. We will assume
that the LED effects can be observed, if\cite{ma}
\begin{eqnarray}
\label{upper} \Delta\sigma =\sigma_{LED}-\sigma_{SM} \geq
\frac{5\sqrt{\sigma_{LED}\mathcal{L}}}{\mathcal{L}},
\end{eqnarray}
where the integrated luminosity ${\mathcal{L}}$ are assumed to be
$500 fb^{-1}$. For $e^+e^-\rightarrow H^+H^-$, from Fig.~5, we
see that the LED can be probed, for example, up to 2800\,GeV and
7200\,GeV with $m_{H^-}=150$\,GeV for $\sqrt{s}=1000$\,GeV and
$3000$\,GeV, respectively. For $e^+e^-\rightarrow H^-t\bar{b}$, from
Fig.~10, we see that the LED can be probed, for example, up to
1400\,GeV and 3600\,GeV for $m_{H^-}=260$ and $1100$\,GeV,  and
$\sqrt{s}=500$, and $2000$\,GeV, respectively.

\section{summary}
In the 2HDM II with large extra dimensions we investigated the
contributions of virtual KK gravitons for the 2HDM II charged
Higgs production in the two important production processes:
$e^+e^-\rightarrow H^+H^-$ and $e^+e^-\rightarrow H^-t\bar{b}$ at
future linear colliders. We found that KK gravitons can
significantly modify these total cross sections and their
differential cross sections compared to the corresponding 2HDM II
values and, therefore, can be used to probe the effective scale
$\Lambda_T$ up to several TeV. For example, at $\sqrt{s}=2$\,TeV
the cross sections for $e^+e^-\rightarrow H^+H^-$ and
$e^+e^-\rightarrow H^-t\bar{b}$ in the 2HDM II are 7.4fb for
$m_{H^-}=150$\,GeV and 0.006fb for $m_{H^-}=1.1$\,TeV with
$\tan\beta=40$, while in LED they are 12.1fb and 0.013fb,
respectively, for $\Lambda_T=4$\,TeV.

\begin{acknowledgments}
We thank Prasanta Kumar Das for useful discussions. This work was
supported in part by the National Natural Science Foundation of
China, under grant Nos.10421003 and 10575001, and the Key Grant
Project of Chinese Ministry of Education, under grant NO.305001,
and the U.S. Department of Energy, Division of High Energy
Physics, under Grant No.DE-FG02-91-ER4086.
\end{acknowledgments}

\begin{figure}[h!]
\vspace{1.0cm} \centerline{\epsfig{file=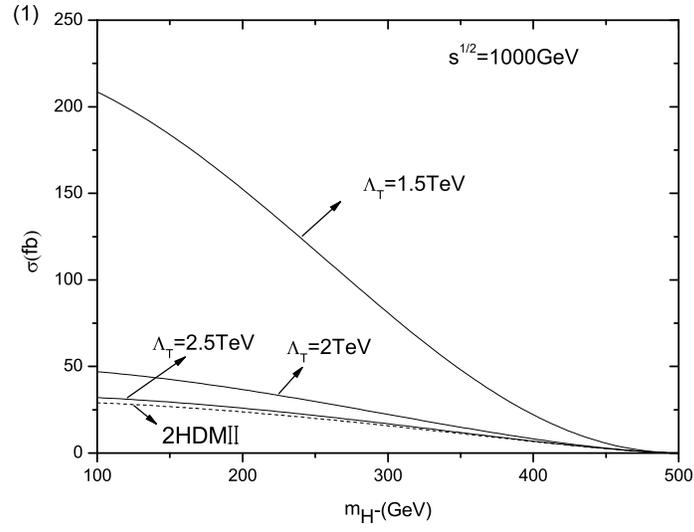, width=300pt}}
\caption[]{Dependence of the cross section in the 2HDM II and LED
for the process $e^+e^-\rightarrow H^+H^-$ on $m_{H^-}$, assuming
$\sqrt{s}=1000$\,GeV.}
\end{figure}
\begin{figure}[h!]
\vspace{1.0cm} \centerline{\epsfig{file=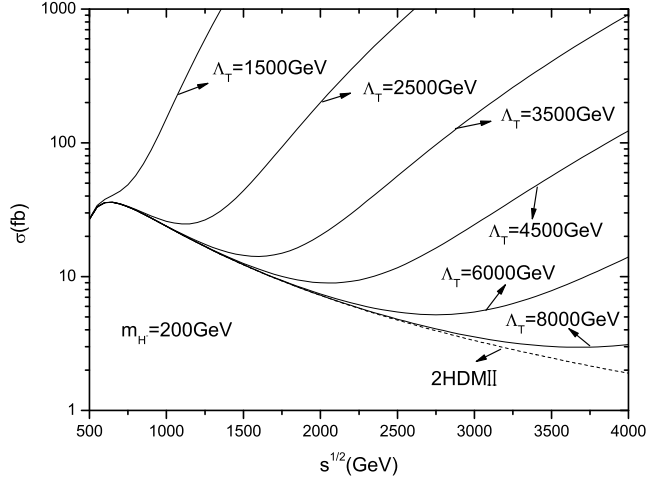, width=280pt}}
\caption[]{Dependence of the cross section in the 2HDM II and LED
for the process $e^+e^-\rightarrow H^+H^-$ on $\sqrt{s}$, assuming
$m_{H^-}=200$\,GeV. }
\end{figure}
\begin{figure}[h!]
\vspace{1.0cm} \centerline{\epsfig{file=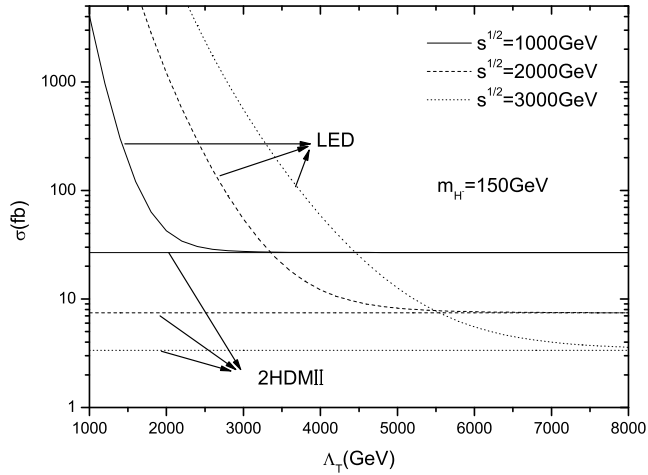, width=280pt}}
\caption[]{Dependence of the cross section in the 2HDM II and LED
for the process $e^+e^-\rightarrow H^+H^-$ on the effective energy
scale $\Lambda_T$, assuming $m_{H^-}=150$\,GeV,
$\sqrt{s}=1000$\,GeV, 2000\,GeV, and 3000\,GeV. }
\end{figure}
\begin{figure}[h!]
\vspace{1.0cm} \centerline{\epsfig{file=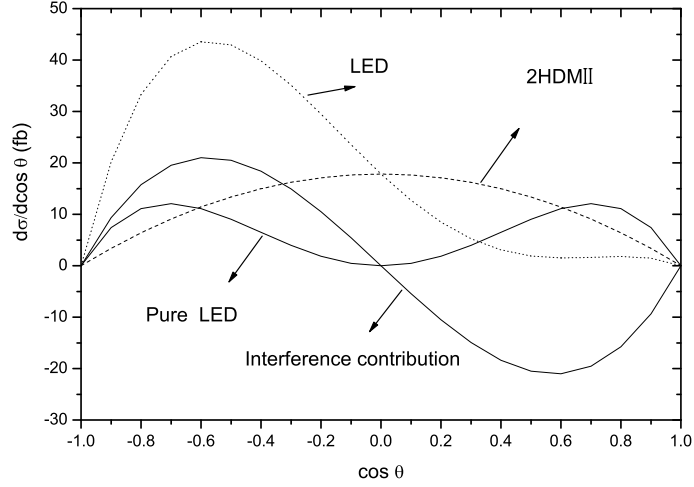, width=300pt}}
\caption[]{The differential cross section $d\sigma/dcos\theta$ in
the 2HDM II and LED for the process $e^+e^-\rightarrow H^+H^-$ as
functions of $\cos\theta$, assuming $m_{H^-}=200$\,GeV,
$\sqrt{s}=1000$\,GeV, $\Lambda_T=2000$\,GeV.}
\end{figure}
\begin{figure}[h!]
\vspace{1.0cm} \centerline{\epsfig{file=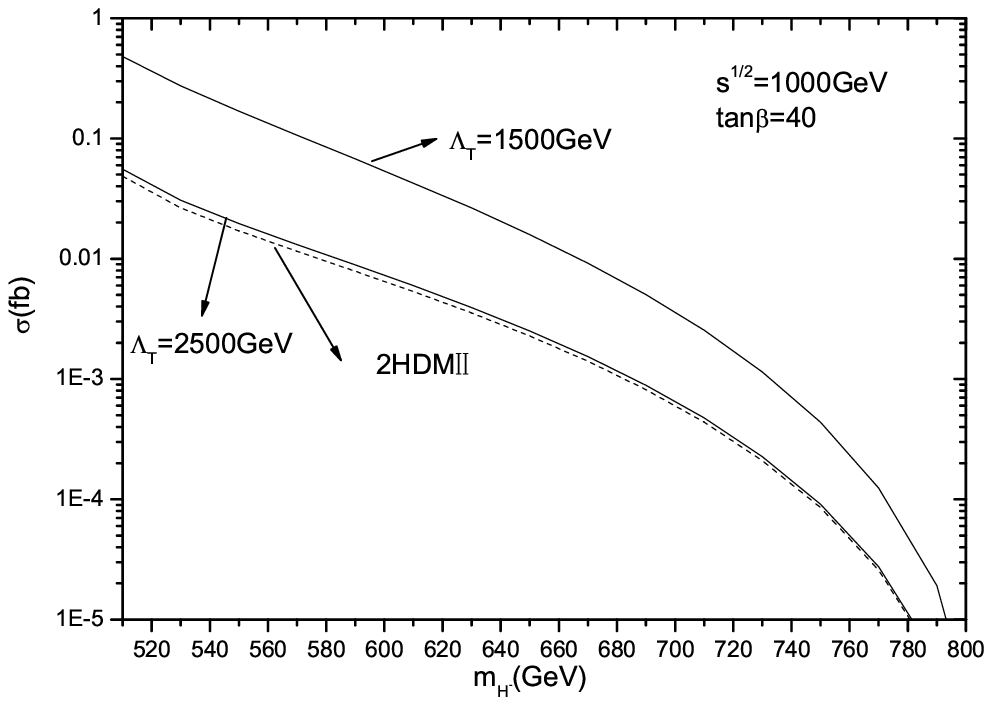, width=300pt}}
\caption[]{Dependence of the cross section in the 2HDM II and LED
for the process $e^+e^-\rightarrow H^-t\bar{b}$ on $m_{H^-}$,
assuming $\sqrt{s}=1000$\,GeV, $\tan\beta=40$. }
\end{figure}
\begin{figure}[h!]
\vspace{1.0cm} \centerline{\epsfig{file=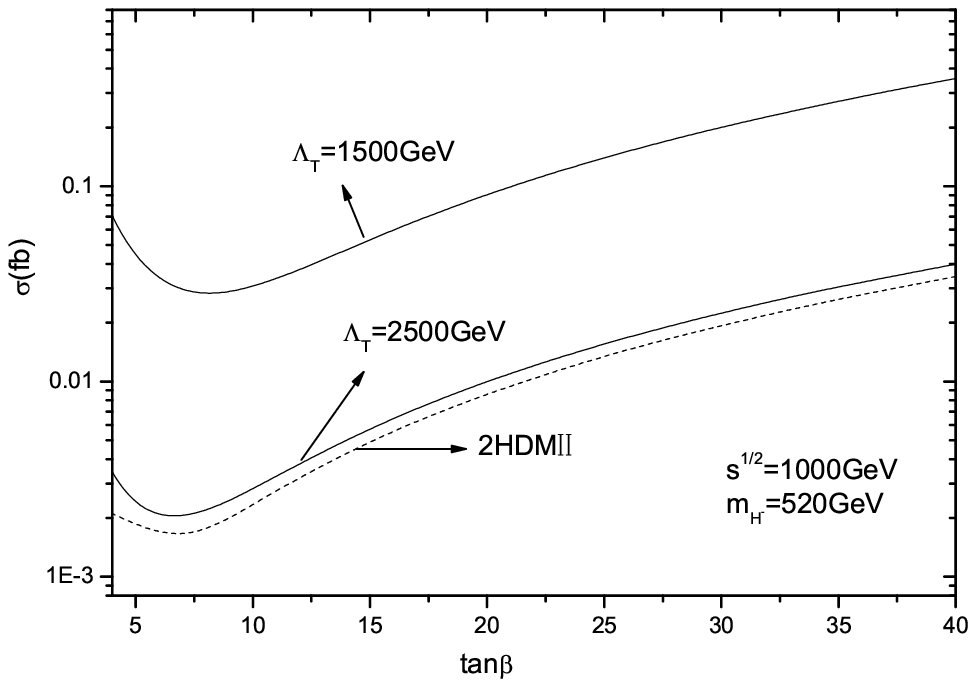, width=300pt}}
\caption[]{Dependence of the cross section in the 2HDM II and LED
for the process $e^+e^-\rightarrow H^-t\bar{b}$ on $\tan\beta$,
assuming $\sqrt{s}=1000$\,GeV, $m_{H^-}=520$\,GeV.}
\end{figure}
\begin{figure}[h!]
\vspace{1.0cm} \centerline{\epsfig{file=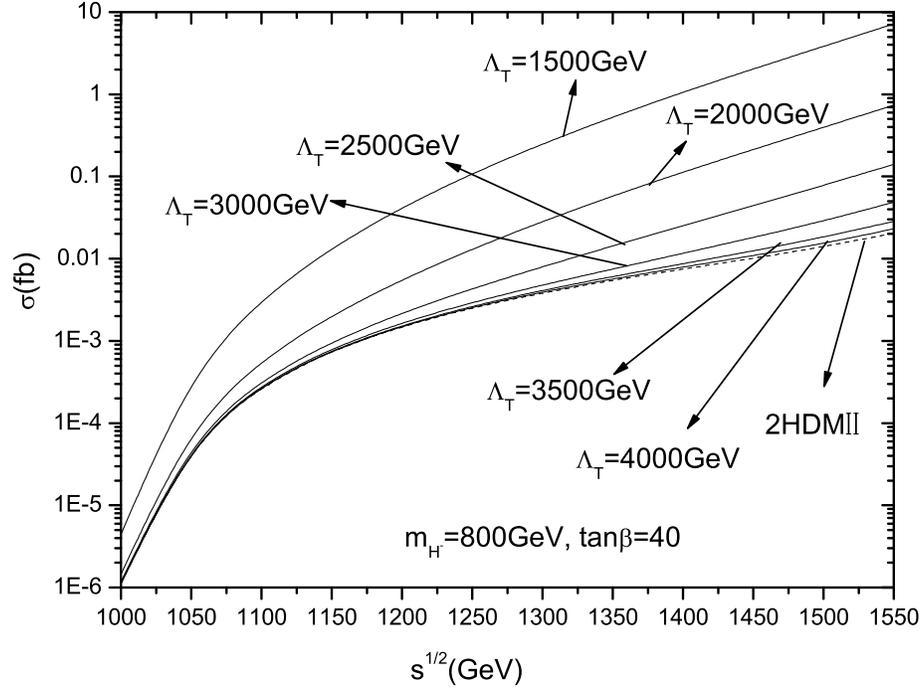, width=400pt}}
\caption[]{Dependence of the cross section in the 2HDM II and LED
for the process $e^+e^-\rightarrow H^-t\bar{b}$ on $\sqrt{s}$,
assuming $m_{H^-}=800$\,GeV, $\tan\beta=40$.}
\end{figure}
\begin{figure}[h!]
\vspace{1.0cm} \centerline{\epsfig{file=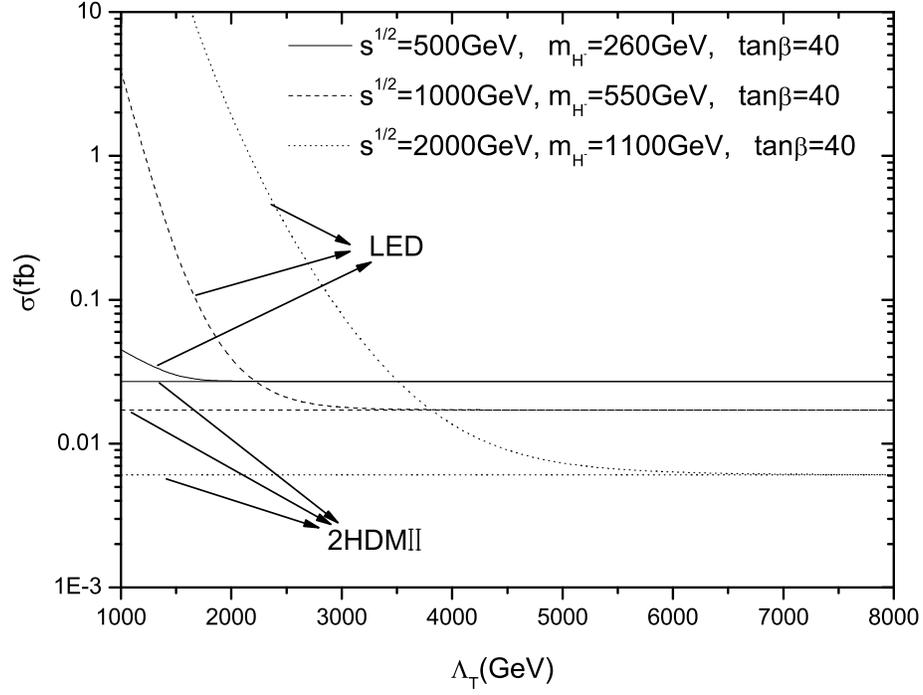, width=400pt}}
\caption[]{Dependence of the cross section in the 2HDM II and LED
for the process $e^+e^-\rightarrow H^-t\bar{b}$ on the effective
energy scale $\Lambda_T$, assuming $m_{H^-}=260, 550, 1100$\,GeV,
and $\sqrt{s}=500, 1000, 2000$\,GeV, respectively.}
\end{figure}
\begin{figure}[h!]
\vspace{1.0cm} \centerline{\epsfig{file=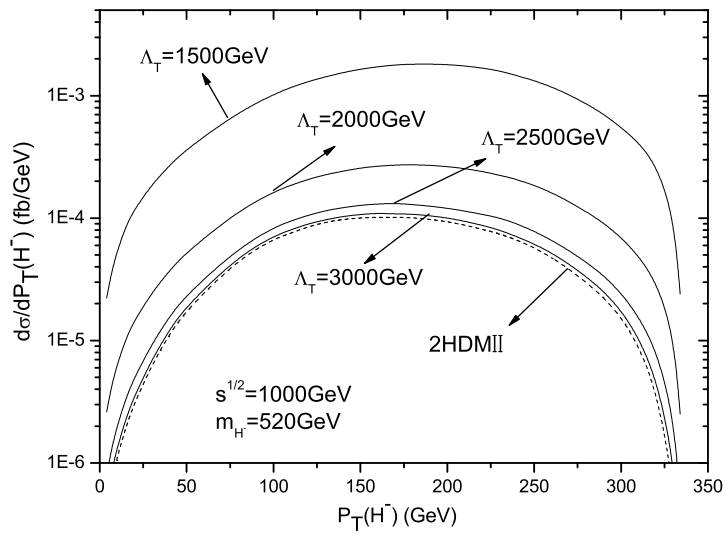, width=320pt}}
\caption[]{The differential cross section $d\sigma/dP_T(H^-)$ in
the 2HDM II and LED for the process $e^+e^-\rightarrow
H^-t\bar{b}$ as a function of the transverse momentum of the
charged Higgs boson $P_T$, assuming $m_{H^-}=520$\,GeV,
$\sqrt{s}=1000$\,GeV, $\Lambda_T=1500$\,GeV, $2000$\,GeV,
$2500$\,GeV and $3000$\,GeV.}
\end{figure}
\end{document}